# Machine Learning - driven insights for predicting the impact of nanoparticles on the functionality of biomolecules, Illustrated by the case of DNA Damage-Inducible Transcript 3 (CHOP) inhibitors

Mariya L. Ivanova[1,*,] ORCID, Michael Nicholls[2,] ORCID, Nicola Russo[1,] ORCID , Gueorgui Mihaylov[3,] ORCID, Konstantin Nikolic[1,] ORCID

Author affiliations
[1]School of Computing and Engineering, University of West London, London, UK
[2] The University of Law, UK
[3]Haleon, London, UK
*Corresponding author mariya.ivanova@uwl.ac.uk

## Abstract

This study introduces a pioneering machine learning (ML)-based approach for predicting the impact of nanoparticle (NP) carriers on the functionality of attached small biomolecules. It was hypothesised that NP interactions induce measurable perturbations in the atomic environment of the small biomolecules, which are reliably captured by chemical shifts in 13C and 1H NMR spectroscopy. Ten datasets were generated by combining 13C, 1H NMR spectroscopy data, derived from SMILES notations and molecular features provided by PubChem. The resulting datasets were used to train predictive models via traditional ML algorithms (Scikit-learn) and Deep Neural Network DNN (PyTorch). The methodology was demonstrated through a quantitative high-throughput screening (qHTS) focused on DNA Damage-Inducible Transcript 3 (CHOP) inhibitors. The optimal ML performance was achieved by the Random Forest Classifier, which was trained on 19,184 samples and tested on 4,000, resulting in 81.1% accuracy, 83.4% precision, 77.7% recall, 80.4% F1-score, 81.1% ROC, and a five-fold cross-validation score of 0.821. Complementing the main study, two computational approaches were developed to enhance CHOP inhibitor prediction. The first identifies the most desirable/undesirable functional groups for CHOP inhibition. The second, a CID_SID ML model, achieved 90.1% accuracy in predicting whether compounds designed for other purposes possess CHOP inhibition potential.

## Introduction

Nanoformulations (NFs), constructed from biomolecules and nanoparticles (NPs) into nanoscale architectures, are designed to augment biomolecule efficacy across various medical applications, such as drug delivery and tissue engineering, theragnostics, imaging, sensing, vaccine development, and medical nanodevices. This convergence of nanotechnology and medicine has spawned the field of nanomedicine. Although a relatively new field, nanomedicine has achieved substantial progress, with approximately 100 nanotherapeutics approved or under FDA review [1].

However, predicting NPs behavior, as a part of such NFs, presents significant challenges. Due to their nanoscale dimensions, NPs cannot be accurately characterized using standard light microscopy [2]. Furthermore, their dimensions approach the quantum realm and hold the potential to alter inherent properties. Brownian forces, particularly influential on sub-micron particles, complicate particle motion control. Inconsistencies observed in particle sizing using various methods emphasize that the size measurement of the NPs is not an authentic feature [3]. Moreover, human serum (HS) can alter NP`s size and surface potential due to albumin adsorption and/or fibrinogen aggregation, as shown in studies with poly(lactic-*co*-glycolic) acid **(**PLG) NPs in buffer saline [4]. Also, the conductance state of NPs is size-dependent. Ultrafast laser spectroscopy has revealed the transitions in gold NPs from metallic to non-metallic behaviour based on size changes [5]. Notably, NPs with similar atom counts and sizes can exhibit significantly different toxic effects [6]. A critical concern regarding NPs prediction extends beyond well-documented issues like cytotoxicity, genotoxicity, immunogenicity, unintended organ accumulation, and long-term side effects [7, 8, 9]. To address these concerns, specialized methodologies have been developed. One such approach is the nano-quantitative structure-activity relationship (nano-QSAR), an adaptation of the well-established QSAR model used in chemistry and pharmacy [10] Nano-QSAR aims to correlate nanoparticle structure with biological activity. While promising, it currently lacks universality and requires further refinement [11]. Similarly, the structure and activity prediction network (SAPNet), designed to guide NP design by identifying structural modifications for desired properties [12], also necessitates improvement. Another QSAR-derived method utilizing simplified molecular input-line entry systems (SMILES) [13] considers the molecular structure and electrical data to predict endpoints, but its NF-specific adaptation remains incomplete, highlighting the need for further research [10]. Beyond QSAR variations, quantum mechanics has been applied to estimate nanoparticle targeted delivery efficiency [14], although the reliance on wireless electromagnetic radiation systems introduces potential sources of error. Proof-of-concept models using iron oxide NPs demonstrated the feasibility of simulating nanomaterial impacts on living organisms with machine learning (ML) [15].

Given the cost and time required to develop medicines, which takes USD 100 million to USD 2 billion over an average period of 10 to 17 years to bring a single drug to market [16], plus the fact that overall, 9 out of 10 drug candidates that have entered clinical trials never submit for FDA approval [17], ML with its ability to predict promising- and fail- drug candidates in the early stage of their development, has the potential to prevent further investment in non-valued formulations and save a significant portion of the development cost per drug.

The current study explored the application of ML and nuclear magnetic resonance (NMR) spectroscopy data for addressing the limitations in NP prediction. For structural analysis, NMR spectroscopy is generally unsuitable for the NP core itself. This limitation arises because the high atomic density and crystalline nature of solid, metallic NP cores cause extreme signal broadening and short relaxation times, often resulting in complete signal loss. The study, however, is not focused on analysing the NP core structure, but rather on predicting the NP impact on the functionality of the attached small biomolecule (drug) over time. It is presumed the NP has already been designed and synthesised. It has been considered that the NMR technique is highly sensitive to changes in the molecular environment, such as variations in pH or molecular conformation. This allows for the accurate measurement of the pKa of ionizable groups or the detection of crucial bonding, key factors that dictate a biomolecule's functional state. So, the spectroscopic data, while not NP-related in terms of training the ML models, serve as a direct surrogate measure for the subtle molecular-level changes in the small biomolecule induced by its conjugation to the NP surface, which information is closely related to the functionality of biomolecules [18]. Electromagnetic radiation absorption by atomic nuclei in a strong magnetic field allows for the exploitation of their magnetic properties. The subsequent relaxation of the excited nuclei, accompanied by the emission of radiation, provides a spectrum of frequencies. These frequencies, or chemical shifts, are highly sensitive to the electronic environment of the nuclei, enabling the elucidation of molecular structure and dynamics. Carbon-13 isotope ($^{13}C$) NMR spectroscopy identifies the chemical environment of carbon atoms within a biomolecule's carbon skeleton [19]. The proton (1H) NMR spectroscopy, on the other hand, identifies different hydrogen nuclei and their magnetic properties, revealing hydrogen bonding patterns that contribute to the understanding of intermolecular interactions [20]. The difference between both types of NMR spectra is that the former is a single peak corresponding to each unique carbon environment and is simpler than the latter, whose complexity is due to the spin-spin coupling between neighbouring protons [21]. So, considering the established dependencies of biomolecule functionalities mentioned above and the elucidative capabilities of NMR, providing information for the chemical environment of the carbon atoms building the carbon “skeleton “of the small biomolecule, it was hypothesised that an ML model trained on NMR spectroscopic data could predict the influence of NPs on biomolecule functionality, testing the NF both before and after it has been in a biological fluid such as blood plasma, cell culture media, or interstitial fluid and thus considered the influence of the “protein corona”.

The necessary NMR spectroscopy data was obtained by converting the small biomolecules` SMILES notations [13] into 13C and 1H NMR spectroscopy data using the NMRDB online tool [22]. This tool employs HOSE code (Hierarchical Organisation of Spherical Environments) methods, and the quantitative errors of the predictions can vary depending on the specific compound [23] In general, the Mean Absolute Error (MAE) of 1H NMR chemical shifts (protons) are in the range of 0.2-0.3 ppm [24], and of 13C NMR chemical shifts (carbons) are around 3ppm [25]. Even though the data used in the study were synthetic with fluctuating errors, they were sufficient to conduct and illustrate a preliminary study of the strategy concept.

In the available literature, NMR spectroscopy has been pointed as a useful addition to electron microscopy and optical absorption spectroscopy used for characterisation of NPs, particularly for the hard–soft matter interfaces [26] and recognised as a technique capable of bridging the analytical gap between NPs in solution and solid phases [27]. NMR techniques have been employed to provide a method for elucidating the morphology and dynamics of polymer-functionalised NPs, with potential application to complex systems that form coronas around NPs [28]. Another study explored quantitative and one– and multi-dimensional NMR spectroscopy on gold NPs and developed a general method for NPs characterisation with NMR spectroscopy [29]. Overall, the direct influence of NPs on the functional activity of the small biomolecules, which they are intended to assist, requires investigation that can be supported by the NMR spectroscopy, covering Chemistry analysis of the NPs, their structural and dynamic characterisation and detection of their interactions with other molecules or materials [30].

The presented approach followed the methodology of two prior studies [31, 32] that predicted human dopamine D1 receptor antagonists and Transthyretin (TTR) transcription activators, respectively. Both studies employed ML algorithms from the scikit-learn library [33]. The ML data was derived from SMILES notations converted to $^{13}C$ NMR spectroscopy features by the NMRDB software. The molecular features of the small biomolecule, pre-calculated by PubChem [34], XLogP3 [ 35], and CACTVS [36] and provided by PubChem, have already shown their potential for ML development [37, 38] and explored in the study focused on predicting Transthyretin transcription activators [32]. These features were:

(i) Molecular weight (MW) as a sum of the mass of all constituent atoms [39].
(ii) Topological Polar Surface Area (TPSA) [40]
(iii) XLogP3-AA (XL), which is a predicted octanol-water partition coefficient [41].
(iv) Hydrogen Bond Donor Count (HBDC) in the given small biomolecule.
(v) Hydrogen Bond Acceptor Count (HDAC) in the given small biomolecule.
(vi) Rotatable Bond Count (RBC). For a bond to be rotatable, it must be a single bond, not part of a ring, and connect two atoms that are not hydrogen and are not at the end of a chain.

The current study expanded upon the methodology of the two preliminary studies by incorporating $^{1}H$ NMR spectroscopy data. The key difference from these preliminary studies was the use of consecutive decimal numbers, rather than natural numbers, to define chemical shift subranges for feature generation, allowing a more in-depth analysis of peak counts. Unlike the preliminary studies, which used traditional ML approaches, this study employed a PyTorch-based deep neural network (DNN) [42] with Optuna-optimized hyperparameters [43]. Python [44] and Jupyter Notebook [45] were employed as the programming language and development environment for all prediction models.

Data derived from PubChem AID 2732 bioassay [46] focused on predicting the C/EBP Homologous Protein (CHOP) inhibitors was used to demonstrate the methodology. CHOP is a crucial transcription factor in the apoptotic arm of the Unfolded Protein Response (UPR). It can be activated by the accumulation of aberrantly folded proteins that have been recognised by the cellular surveillance system and retained within the endoplasmic reticulum (ER) [47,48]. The transcription factor activates ER protein chaperones and mediates for UPR response. So, it has been hypothesised that the inhibition of CHOP could regulate the unfolded protein response to ER stress and would have a potential therapeutic application to diverse diseases [49, 50, 51], such as diabetes [52], Alzheimer's disease [53] (although, it has been reported that CHOP is not the primary contributor to tau-mediated toxicity, related to memory lost [54]), Parkinson's disease [55], haemophilia [56], lysosomal storage diseases [57], and alpha-1 antitrypsin deficiency [58].

The paper also presents two computational applications for CHOP inhibition, in addition to the main research. The first one ordered the functional groups/fragments of explored small biomolecule from the most to the least desirable with respect to the CHOP inhibition. This computational approach was based on data encoded in the chemical names derived by the International Union of Pure and Applied Chemistry (IUPAC) nomenclature, which by itself ensure unique names for any chemical compound. The methodology has been developed and demonstrated with a case study on Tyrosyl-DNA phosphodiesterase 1 (TDP1) inhibitors [59]. Through developing this approach with CHOP-related data, the results serve researchers interested in CHOP inhibition and explore the applicability of the methodology for different data than TDP1. The second computational approach developed beyond the main study was the CID_SID ML model that can predict CHOP inhibitors, using only PubChem identifiers, i.e. PubChem CID and PubChem SID. This approach enables the assessment of small biomolecules initially intended for other targets for their CHOP inhibition capability. Since, generally, the identifiers are not used for ML training and testing, the CID_SID ML model is unconventional. Despite this, its development was meaningful because PubChem's method considers structural and similarity data when generating their identifiers. [60]. The CID_SID ML model has already been developed and is available in the relevant study [61].

## Methodology

The methodology is illustrated in Figure 1 and Figure 2. The columns with CIDs, SMILES notations and activity labels of small biomolecules were retrieved from the PubChem AID 2732 dataset [46]. Since the study was focused on classification ML models, the severe imbalance between the inactive and active small biomolecules was handled. For this purpose, the inactive samples were reduced by keeping only the compounds considered as well in PubChem AID 1996 bioassay focused on small biomolecule solubility [62]. After shuffling the reminded inactive compounds, each second sample was selected and kept. Thus, the inactive compounds were reduced to some extent and combined with all active compounds from the PubChem AID 2732 dataset [46].

From the resulting dataset, the following datasets were created:

(i) A dataset, which included only CIDs and SMILES notations, was formulated to enable the acquisition of NMR spectroscopic data through the NMRDB software.
(ii) A dataset containing only SMILES notations and the activity labels of the small molecule was used later for labelling the spectroscopic data
(iii) A dataset containing only CIDs, which was used as a list for downloading the molecule features listed above of the small molecules.

Once the spectroscopic chemical shifts were obtained for 1H NMR and 13C NMR spectroscopy data, two types of datasets were generated. In the first type, each pick along the chemical shifts scale was counted within subranges defined by consecutive integers and called concise. The subranges in the second type were defined by consecutive decimal numbers and called extensive. All subranges formed the newly generated features of the data frames, which contained the number of picks along the chemical shifts scale. The resulting four datasets, containing concise 1H NMR, concise 13C NMR, extensive 1H and extensive 13 C NMR spectroscopy data, were combined, as illustrated in Table 1, and eight datasets were obtained. Each of these eight datasets was used further for ML with the classifiers: Decision Tree, Random Forest, Support Vector and Gradient Boosting software interpreted by scikit-learn ML library [35]. ML was conducted based on the best practices recommended in the literature [63, 64]. For that purpose, an equal number of samples for each class were extracted to ensure that there would not be bias towards the majority class that would lead to misleading accuracy. The remaining samples were then balanced, increasing the number of minor classes with randomly selected and repeated samples of this class until the number of samples in the minor class was equal to the number of samples in the major class. The ML models were then conducted with each of these eight datasets. The ML metrics were compared, and the most suitable dataset and an ML classifier for this case study were selected and scrutinised for overfitting tracing the deviation between the training and testing accuracy to be lower than 5%.

Further, the above-described molecular features were integrated into the dataset used by the optimal ML model. Following this, ML analyses were conducted, comparing the results with and without PCA application [65], which application reduced the number of features, and the results obtained using only the molecular features. Using the expanded dataset with molecular features, a PyTorch DNN was developed, with hyperparameters optimised by Optuna [43] and scrutinised for overfitting. DNN was trained ten times, and the mean accuracy was compared to the optimal scikit-learn ML model.

The supplementary computational approaches followed the methodologies of the relevant studies. About this one using data encoded in the IUPAC names [59], the IUPAC names of the small biomolecule from the PubChem AID 2732 bioassay` dataset [46] were parsed into strings of four or more letters, and then used in two ways: to develop an ML model and to extract the most desirable and non-desirable functional groups/fragment regarding the CHOP inhibition. For the second ML approach, the CID_SID ML model, the CIDs, SIDs and targets from the PubChem AID 2732 bioassay` dataset [46] were extracted and used for ML [61]. For more details about the methodologies, please refer to the relevant studies [59, 61].

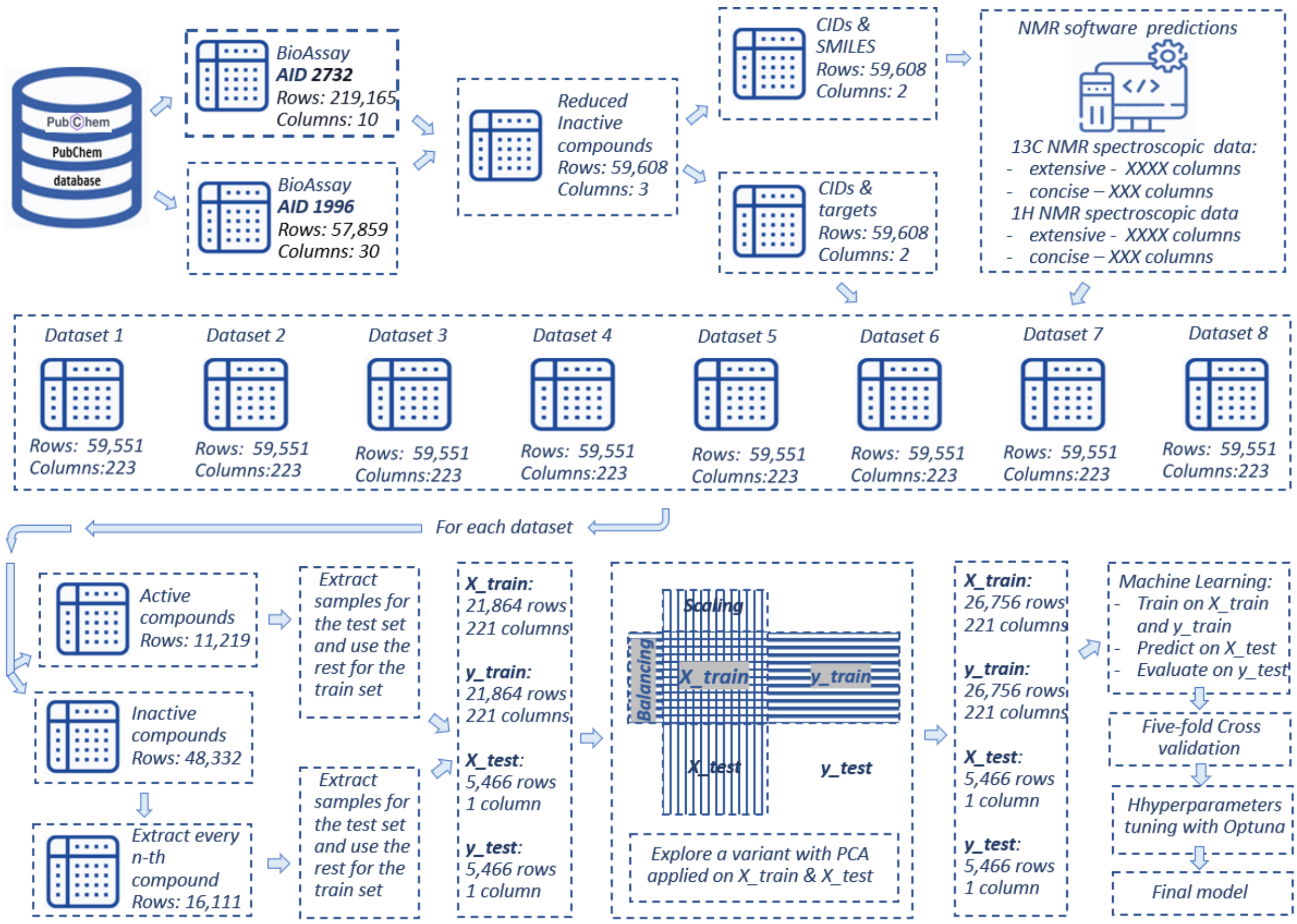

Figure 1 Methodology of ML with eight datasets

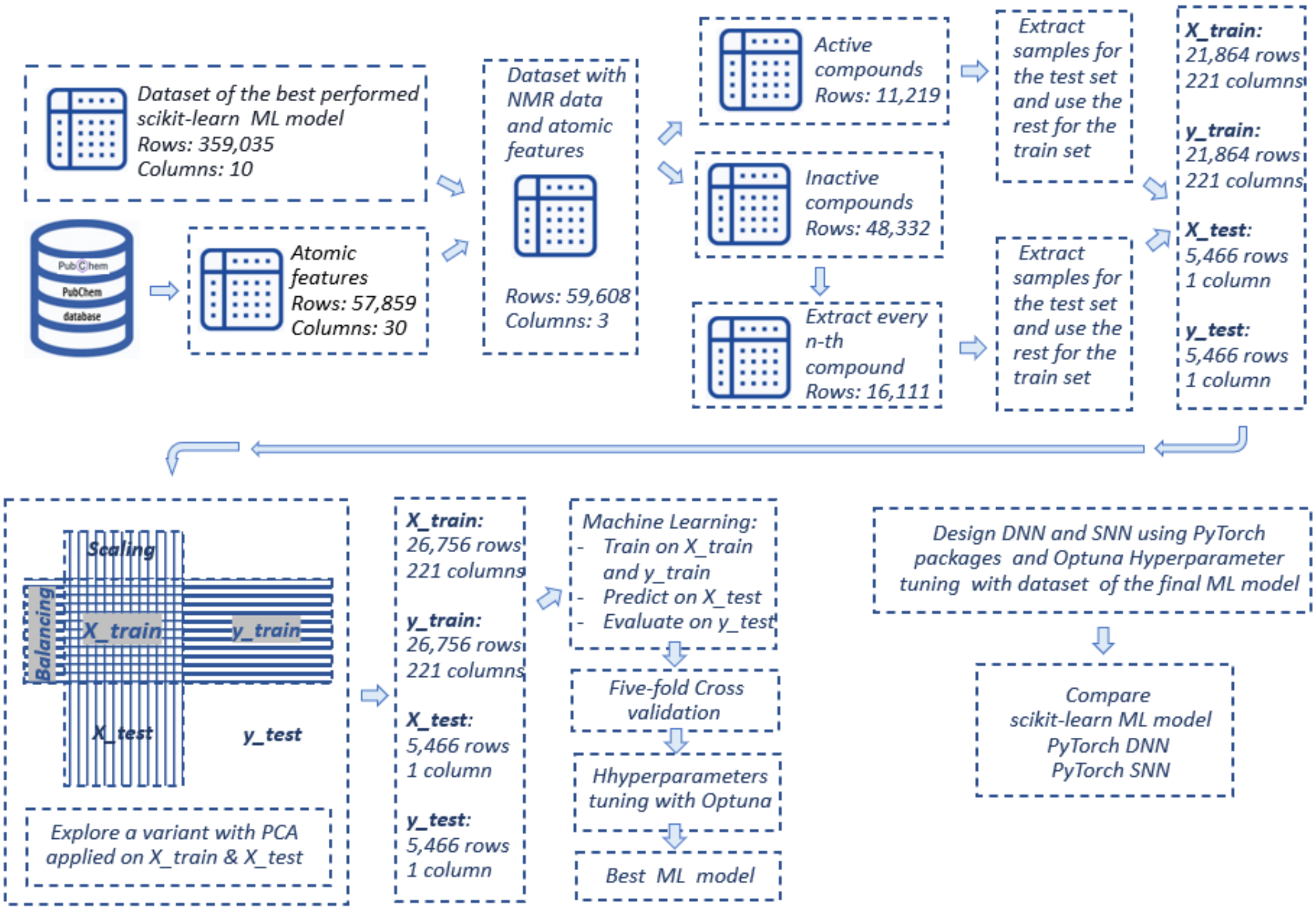

Figure 2 ML performed with the dataset of the optimal ML model and molecular features

## Results and discussion

By identifying the overlap between 210,922 inactive compounds from PubChem AID 2732 and 57,859 samples from PubChem AID 1996, 24,185 inactive samples were retained. The number of these compounds was reduced subsequently, keeping every second sample, which decreased them to 12,085. Combining these 12,085 remaining inactive samples with the 8,224 active compounds from PubChem AID 2732 resulted in a dataset of 20,309 samples. The SMILES notations from this dataset were used by the NMRDB software, and NMR spectroscopy data was obtained as follows:

(i) 220 columns with counts of chemical shifts of 13C NMR spectra in subranges defined by consequent natural numbers; the dataset abbreviation is 13C_c.
(ii) 12 columns with counts of chemical shifts of 1H NMR spectra in subranges defined by consequent natural numbers; the dataset abbreviation is 1H_c.
(iii) 1,828 columns with counts of chemical shifts of 13C NMR spectra in subranges defined by consequent natural numbers; the dataset abbreviation is 13_e.
(iv) 103 columns with counts of chemical shifts of 1H NMR spectra in subranges defined by consequent natural numbers; the dataset abbreviation is 1H_e.

These four datasets, containing NMR spectroscopy data, were combined as detailed in Table 1, resulting in eight datasets subsequently used for ML. To ensure reliable metric results, 2,100 samples were randomly selected from each dataset, resulting in a total of 4,200 samples used for testing the ML models. The imbalance in the remaining compounds was handled using Random Over Sampling (ROS), a technique replicating random minority class samples until both classes contained an equal number of samples. This process yielded a balanced dataset of 16,109 samples, which was used for training the ML models. Since there was a risk of the minority and majority classes not being well-separated, which could lead to synthetic samples blurring the decision boundary, ROS was chosen over Synthetic Minority Over-sampling Technique (SMOTE). Additionally, ROS was preferred over Random Undersampling (RUS) to avoid decreasing the overall data volume, which is crucial for ML model performance.

Table 1 summarises the optimal ML models' accuracy and five-fold cross-validation scores across the eight datasets. The performance of the ML models, when incorporating 13C NMR data, ranged from 76.4% (Dataset 6) to 79.3% (Dataset 2), a difference that was not statistically significant. Furthermore, including 1H NMR data did not improve model accuracy; in fact, a slight decrease was observed. Models trained solely on 1H NMR data yielded the lowest accuracies across the eight datasets, with results of 67.7% (Dataset 3) and 69.5% (Dataset 4). The large gap between single result accuracy and five-fold cross-validation scores indicated potential overfitting. Due to this pattern across all datasets, only the optimal ML model (SVC, Dataset 2) was scrutinized for overfitting.

Initially, SVC based on Dataset 2 (extensive 13 C NMR spectroscopy data) achieved 79.3% accuracy, 82.6% precision, 74.1% recall, 78.2% F1-score, 79.3% ROC (Table ESM 3), and a 0.835 five-fold cross-validation score (standard deviation ±0.002) (Table ESM4) followed closely by SVC based on Dataset 8 (extensive 13 C NMR and extensive 1 H NMR spectroscopy data) 79.2% accuracy, 82.4% precision, 74.2% recall, 78.1% F1-score, 79.2% ROC (Table ESM15) and a 0.844 five-fold cross-validation score (standard deviation ±0.002) (Table ESM16). The SVC variant trained on Dataset 2 was selected to minimize training and testing time due to its lower feature dimensionality. On the other hand, considering both variants, five-fold cross-validation revealed that RFC consistently had the highest cross-validation score, thus ranking it first. However, the substantial difference in RFC performance between a single evaluation and five-fold cross-validation suggests potential overfitting (Table ESM 3, Table ESM4, Table ESM15, Table ESM16). Initial analysis showed no significant overfitting in the SVC model (with default hyperparameter based on Dataset 2) with training and testing accuracies of 0.846 and 0.793, respectively. Optuna's five-trial hyperparameter optimization (C=218090.43, gamma=0.0518) resulted in no significant accuracy gain. However, it increased the training-testing accuracy delta (0.879 vs. 0.796), implying a higher propensity for overfitting.

The inclusion of molecular features in Dataset 2 altered its dimensions, increasing the number of features to 1,934 and reducing the number of samples to 19,501. The optimal ML model performed with this dataset was RFC achieving 83% accuracy, 88% precision, 76.4% recall, 81.8% F1-score, 83% ROC (Table ESM17), and a 0.84 five-fold cross-validation score with 0.006 standard deviations (Table ESM18). The overfitting assessment indicated that max_depth values exceeding 13, where the training and testing accuracy were 84.6% vs. 79.8%, respectively, resulted in a training-testing accuracy deviation greater than 5%, implying a potential for overfitting (Figure ESM 1). The hyperparameter tuning of RFC with Optuna improved the performance of the ML model to 81.1% accuracy, 83.4% precision, 77.7% recall, 80.4% F1-score, 81.1% ROC and 0.821 five-fold cross-validation score. Overfitting (below max_depth=9) was not indicated by the examination (Figure 3). The hyperparameter values suggested by Optuna were:

(i) max_depth=9, define the level the tree can have.
(ii) n_estimators=494, shows the number of trees in the forest.
(iii) min_samples_split=2, the minimum number of samples required to split an internal node.

(iv) min_samples_leaf=6, the minimum number of samples required to be at a leaf node.
(v) max_features=None, i.e. max_features=n_features.
(vi) criterion='entropy', measuring the quality of a split.

The learning curve of the ML model is plotted in Figure 4, the confusion matric in Figure 5, the AUC in Figure 6, and the classification report in Table 2.

Dataset 2, with added molecular features, was used for ML after PCA reduced its dimensionality to the optimal eight components. However, this did not yield performance gains compared to the model without PCA (Table ESM19, Table ESM20). An ML variant using only molecular features was explored, but it did not outperform RFC with molecular-feature-integrated Dataset 2 as well (Table ESM21, Table ESM22).

The Optuna-optimized DNN [43] comprised two hidden layers with 113 and 104 units, respectively. Dropout rates of 0.33165 and 0.3692 were applied after each layer. Utilizing the RMSprop optimizer with a learning rate of 0.001329, the model achieved 82.34% accuracy. However, a significant discrepancy between the final training loss (0.07) and validation loss (2.5), as depicted in Figure ESM2, suggested potential overfitting.

Table 2. Content of the eight datasets which were result of the combination of NMR spectroscopy data; Accuracy of the best ML model with the given dataset before to be scrutinized for overfitting; five-fold cross-validation score of the respected ML model; references to the tables in Electronic Supplementary material (ESM) with full set of ML metrics. Datasets contain data of carbon-13 isotope concise (13C_c), carbon-13 isotope extensive (13C_e), proton isotope concise (1H_c), proton isotope extensive (1H_c),

| Dataset | 13C_c | 13C_e | 1H_c | 1H_e | Accuracy | c-v score | Table |
|---|---|---|---|---|---|---|---|
| 1 | ✓ | - | - | - | 77.9% | 0.864 | ESM1, ESM2 |
| 2 | - | ✓ | - | - | 79.3% | 0.864 | ESM3, ESM4 |
| 3 | - | - | ✓ | - | 67.7% | 0.769 | ESM5, ESM6 |
| 4 | - | - | - | ✓ | 69.5% | 0.821 | ESM7, ESM8 |
| 5 | ✓ | - | ✓ | - | 77.8% | 0.861 | ESM9, ESM10 |
| 6 | - | ✓ | ✓ | - | 76.4% | 0.872 | ESM11, ESM12 |
| 7 | ✓ | - | - | ✓ | 77.8% | 0.861 | ESM13, ESM14 |
| 8 | - | ✓ | - | ✓ | 79.2% | 0.861 | ESM15, ESM16 |
| Dataset 2 & molecule features | | | | | 83.0% | 0.840 | ESM17, ESM18 |

| | | | |
|---|---|---|---|
| Dataset 2 & molecule features; feature reduction with PCA | 82.8% | 0.902 | ESM19, ESM20 |
| Dataset only with molecule features | 80.7% | 0.887 | ESM21, ESM22 |
| PyTorch DNN with Dataset 2 & molecule features | 81.3% | - | Figure ESM2 |

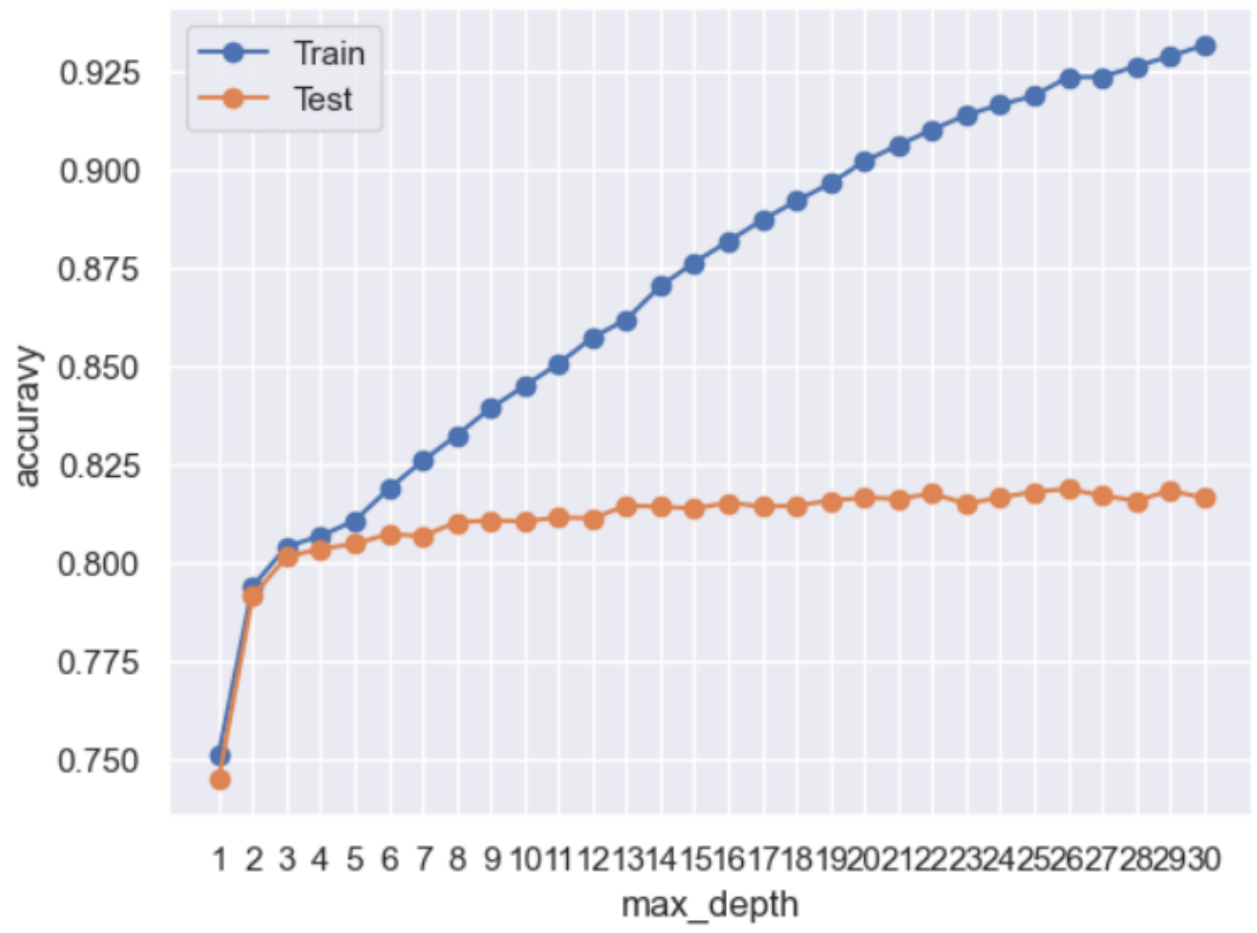

Figure 3 Scrutinising for overfitting of the Optuna-hyperparameter tuned RFC based on Dataset 2 and molecular features

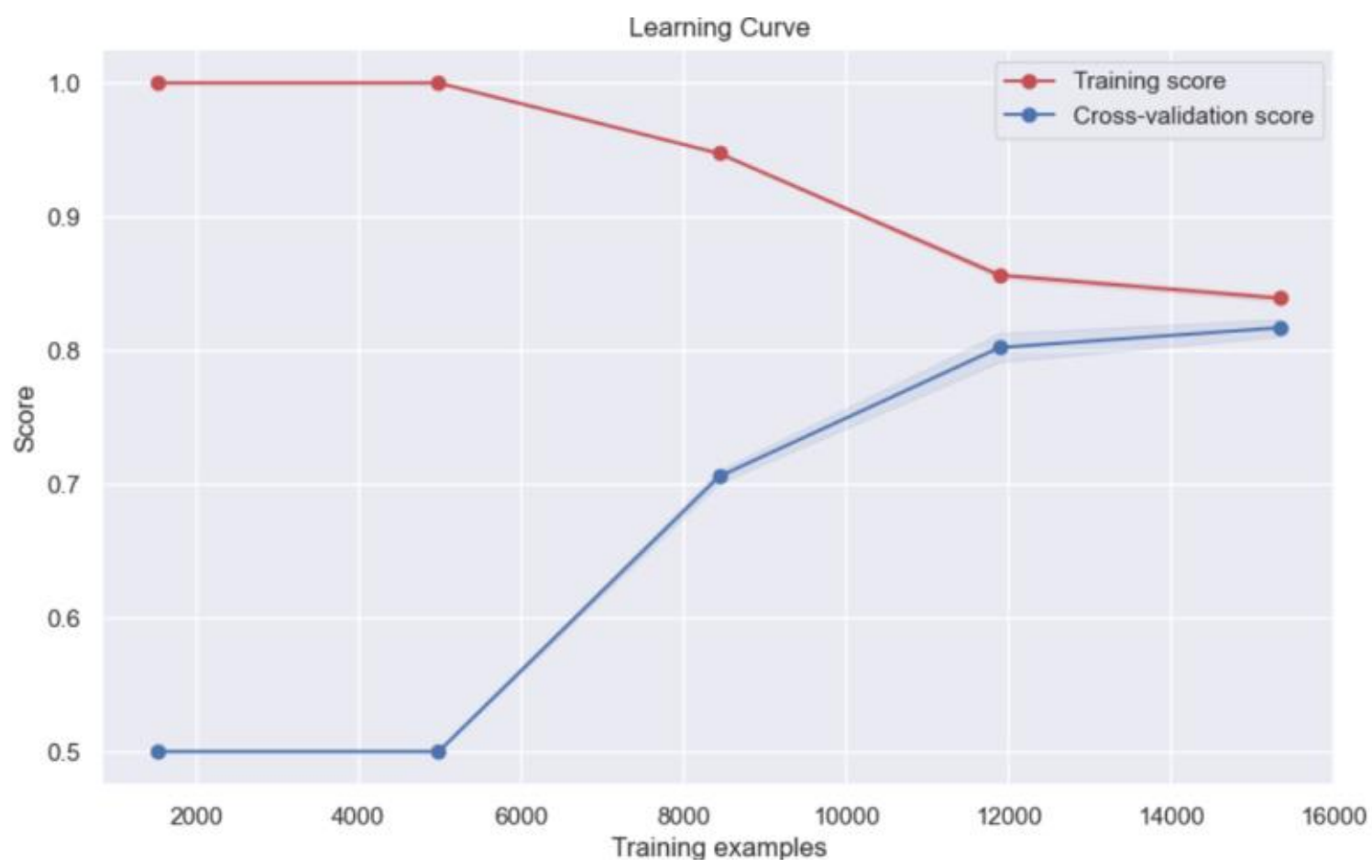

Figure 4 Learning curve of the Optuna-hyperparameter tuned RFC based on Dataset 2 and molecular features

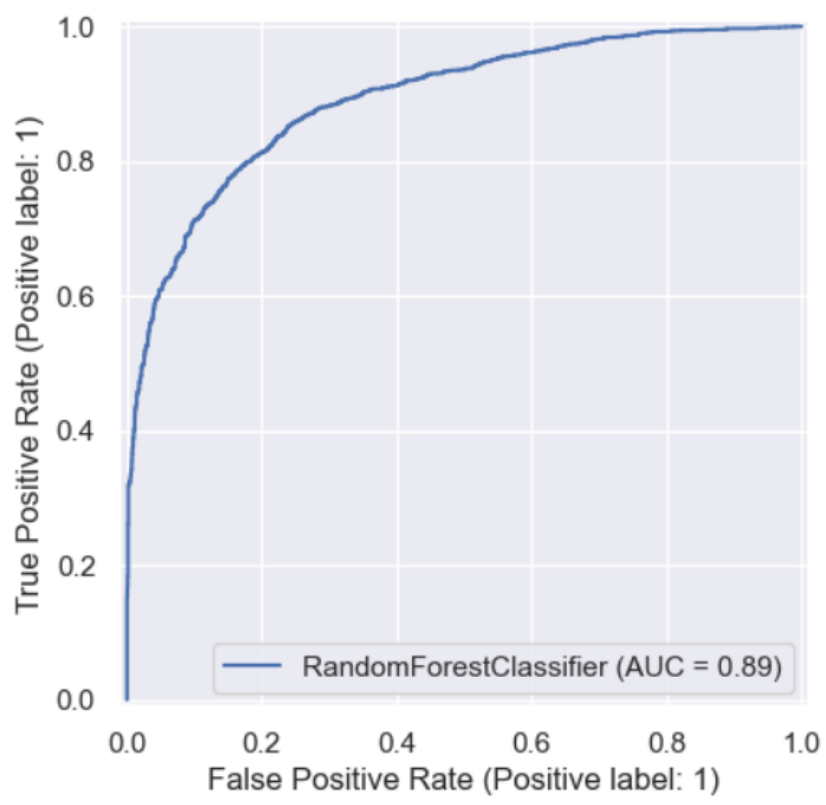

Figure 5. ROC of the Optuna-hyperparameter tuned RFC based on Dataset 2 and molecular features

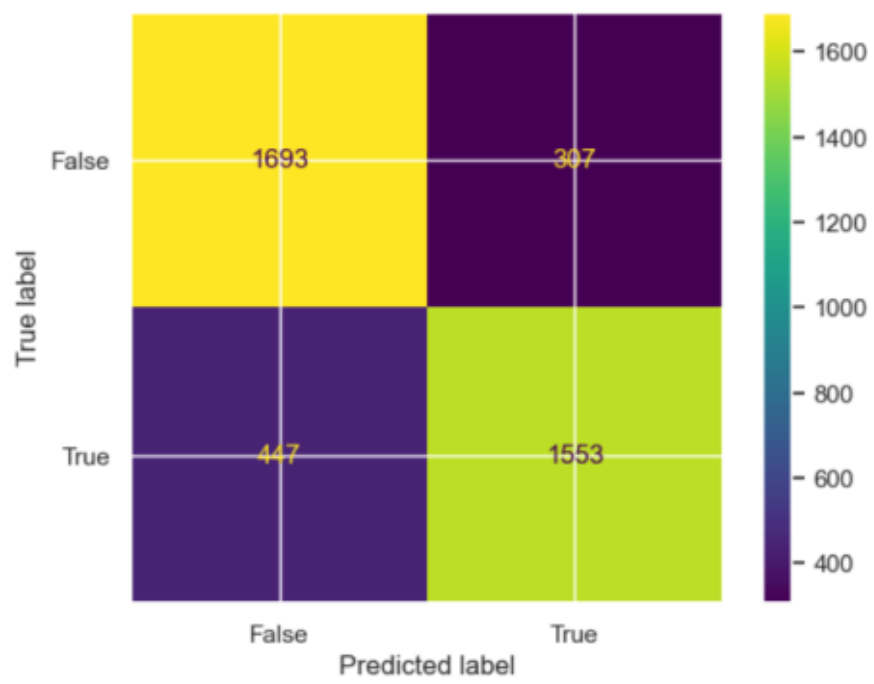

Figure 6. Confusion matrix of the Optuna-hyperparameter tuned RFC based on Dataset 2 and molecular features

Table 2. Classification report of the Optuna-hyperparameter tuned RFC based on Dataset 2 and molecular features

| | precision | recall | F1-score | support |
|---|---|---|---|---|
| Active (target 1) | 0.79 | 0.85 | 0.82 | 2000 |
| Inactive (target 0) | 0.83 | 0.78 | 0.80 | 2000 |
| | | | | |
| accuracy | | | 0.81 | 4000 |
| macro avg | 0.81 | 0.81 | 0.81 | 4000 |
| Weighted avg | 0.81 | 0.81 | 0.81 | 4000 |

The Optuna-tuned RFC achieving 68.9% accuracy, 71.6% precision, 62.5% recall, 66.7% F1-score, 68.8% ROC AUC, and a five-fold cross-validation score of 0.724% (standard deviation ±0.0076). The optimal hyperparameters, determined by Optuna were min_samples_split=10, min_samples_leaf=1, max_features='log2', criterion='gini', and max_depth=15, which was set to prevent overfitting (train/test accuracy deviation < 10%). Figure ESM 3 illustrates the scrutinising for overfitting of the model with default hyperparameters, where the overfitting started at max_depth = 19, where the train accuracy was 66.2%; Figure ESM 4 is a plot of the scrutinising for overfitting of the model hyperparameter tuned by Optuna, which was the final model; Figure ESM 5 confusion matrix and Table ESM 23 the classification report of the final model.

The top 24 functional groups (Figure ESM 6, Figure ESM 7), as ranked by the feature importance and permutation importance algorithms, showed high similarity. The difference between them was three groups per list, namely amino, pyramidin, methyl, sulfamoyl, carbamothioyl, trichloro. The result of combination of two feature importance lists keeping only the unique functional groups/fragments was reordered according to the relevant proportion of the active cases, so with the highest value was sulfamoyl with 69 active and 13 inactive cases and relative proportion of 5.31 and the lowest value relative proportion was hold by carbohydrazide with 0.1 relative proportion of the active cases. The full reordered list based on relative proportion of the active cases is available on GitHub [66]. It should be noted that sulfamoyl should participate in the small biomolecule composition as an independent functional group, not in a pact with other groups, for example like sulfamoylbenzoicacid because it turned out that sulfamoyl name participated in 26 features and formed dataset of 19,499 compounds, where 11,593 are inactive and 7,906 are active, so just the presence of sulfamoyl in the IUPAC name is not enough to conclude that there is a high probability that the compound is a CHOP inhibitor. The sulfamoyl functional group should not be in a composition with other functional groups.

The entire dataset of the bioassay PubChem AID 2732 [46] was used for generation of the lists with functional groups/fragments participate only in active cases (highly desirable) or inactive cases (highly undesirable). As it was mentioned above, the dataset contained 218,583 samples 8,224 of which were labelled as active and 209,952 as inactive. So, the list with highly desirable functional groups/fragments was with 267 elements, starting with hexahydroazuleno and dioxonaphthalene as the most desirable each one of them with five active cases and no one inactive, followed by ynamide, oxirane, pyrazolidine each of them with four active cases The full list is available on GitHub [67]. It is worth mentioning that the hypothesis proposed in the parent study of this approach [59] suggested that when the tested compound contains a functional group from one of these two lists and its (of the tested compound) 13CNMR spectroscopy data is similar to the compound source of this functional group/fragment, there is a high probability that the tested compound is a CHOP inhibitor. Given the nature of 13C NMR spectroscopy, the hypothesis was expected to be applicable in the presence of NPs (i.e. for NFs) as well. In this case the NFs will be processed in the same manner explained in the main study, i.e. way explained in the main study, i.e. NFs to be stood in a complex biological fluid in order to provoke NPs aggregation and then to process NMR spectroscopy whose data will be used for comparison.

Regarding the most undesirable functional groups/fragments, i.e. whose compounds participate only in inactive cases, the descendent order contained 7,155 rows on the top of which was methyl with 25,955 inactive cases followed by carboxamide with 18,520 and methoxyphenyl with 13,153. The list with the first 3000 rows is available on GitHub [68].

Regarding CID_SID ML model that check if a compound, designed initially for other purpose different than CHOP inhibition is a CHOP inhibitor, the RFC obtained 90.1% accuracy, 98.3% precision, 81.7% recall, 89.2% F1, 90.1% ROC, five-fold cross-validation score of 0.943 with standard deviation of ±0.00075. For more details, please refer to the original study [61]

Given the highly **structured and empirical nature** of the presented methodology, the obtained results are fundamentally **free of** AI **hallucinations**. This reliability is ensured because the ML models are not generative; instead, their inputs are strictly **N**MR **chemical shifts**, features derived from the SMILES notation of known, chemically sound molecules and processed by the NMRDB tool. The models are trained only to find correlations between these verifiable spectroscopic data and biological activity, preventing the random feature generation or structural invention characteristic of generative AI hallucination. [69, 70].

## Conclusion

An ML methodology was developed, leveraging molecular features and 1H and 13C NMR spectroscopy data derived from SMILES notations. It was hypothesised that the approach could be applicable in case of predictions of a NF functionality due to the capability of 13C NMR to provide information regarding the chemical environment of the carbons building the carbon skeleton of small biomolecules. So, the presence of NP with or without protein corona would affect this environment and respectively detected by the 13C NMR spectroscopy. This information in turn can be used by ML to find a pattern amongst data that can predict the functionality of the compound (NF) accommodating the NP. The methodology was demonstrated using CHOP inhibitors as a case study, but its applicability to other bioassays is anticipated. A key innovation of this research, within a broader investigation, was the refined segmentation of chemical shift ranges, which increased feature dimensionality and slightly enhanced ML model performance. Additionally, a PyTorch DNN was designed and optimised using Optuna. The optimal solution was RFC that scrutinised for overfitting to confirm its robustness. Although data used for the development of the ML model was synthetic, the NMR chemical shift of NF that will be tested must be real spectroscopy data. The complementary computational approaches provide the researchers interested in CHOP inhibition with insights related to drug discovery and a side effect. The computational methodology was not experimentally validated in this study due to the lack of necessary laboratory and NMR spectroscopy facilities. The approach remains open for further investigation by interested researchers with access to the required testing resources.

## Scientific contributions

- Proposed a new concept to address the challenges in predicting the influence of NPs based on 13C NMR spectroscopy.
- Developed an ML model for predicting CHOP inhibition based on SMILES notations.
- Sulfamoyl was identified as having potential to contribute to CHOP inhibition
- Hexahydroazuleno and dioxonaphthalene were identified as a highly desirable functional group for CHOP inhibition.
- Methyl was found to be a highly undesirable functional group for CHOP inhibition
- Developed CID_SID ML model that leverages PubChem CID and SID to predict CHOP inhibition as a potential side effect for existing chemical compounds.

## Author Contributions

MLI, MN, NR, GM and KN conceptualized the project and designed the methodology. MLI and NR wrote the code. MLI, NR and GM processed the data. KN supervised the project. All authors were involved with the writing of the paper.

## Acknowledge

MLI thanks the UWL Vice-Chancellor's Scholarship Scheme for their generous support. We sincerely thank NIH for providing access to their PubChem database. Article is dedicated to Luben Ivanov.

## Data and Code Availability Statement

The Python code is available in Jupyter notebooks files on GitHub:
https://github.com/articlesmli/13C_NMR_ML_model_CHOP.git

Datasets generated during the study are available on Hugging face:

- CID_SID ML model [71]
- molecular features [72]
- IUPAC approach [73]
- 13C NMR [74, 75]
- 1H NMR [76, 77]

The raw data used in the study is available through the PubChem portal:
https://pubchem.ncbi.nlm.nih.gov/

## Conflicts of Interest

The authors declare no conflict of interest.

# Electronic Supplementary Material

## Tables

Table ESM1 Metrics of ML models based on Dataset 1 (13C NMR concise data)

| 1.Algorithm | 2.Accuracy | 3.Precision | 4.Recall | 5.F1 | 6.ROC |
|---|---|---|---|---|---|
| SVM | 0.779 | 0.798 | 0.747 | 0.772 | 0.779 |
| RandomForest | 0.713 | 0.842 | 0.524 | 0.646 | 0.713 |
| GradientBoost | 0.706 | 0.716 | 0.685 | 0.700 | 0.706 |
| Decision | 0.634 | 0.663 | 0.544 | 0.598 | 0.634 |
| K-nearest | 0.634 | 0.760 | 0.392 | 0.517 | 0.634 |

Table ESM2. Five-fold cross-validation for ML based on Dataset 1 (13C NMR concise data)

| 1.Algorithm | 2.Mean CV Score | 3.Standard Deviation | 4.List of CV Scores |
|---|---|---|---|
| RandomForest | 0.8642 | 0.0470 | [0.8331, 0.8294, 0.8161, 0.9189, 0.9237] |
| SVM | 0.8127 | 0.0122 | [0.8129, 0.8019, 0.7975, 0.8311, 0.8203] |
| Decision | 0.7686 | 0.0528 | [0.7351, 0.7238, 0.7187, 0.8263, 0.8391] |
| K-nearest | 0.7386 | 0.0399 | [0.7085, 0.7143, 0.6964, 0.783, 0.7905] |
| GradientBoost | 0.7223 | 0.0072 | [0.7295, 0.7238, 0.709, 0.7272, 0.722] |

Table ESM3 Metrics of ML models based on Dataset 2 (13C NMR extensive data)

| 1.Algorithm | 2.Accuracy | 3.Precision | 4.Recall | 5.F1 | 6.ROC |
|---|---|---|---|---|---|
| SVM | 0.793 | 0.826 | 0.741 | 0.782 | 0.793 |
| RandomForest | 0.711 | 0.854 | 0.510 | 0.639 | 0.711 |
| GradientBoost | 0.673 | 0.696 | 0.613 | 0.652 | 0.673 |
| Decision | 0.653 | 0.683 | 0.572 | 0.623 | 0.653 |
| K-nearest | 0.644 | 0.789 | 0.392 | 0.524 | 0.644 |

Table ESM4. Five-fold cross-validation for ML based on Dataset 2 (13C NMR extensive data)

| 1.Algorithm | 2.Mean CV Score | 3.Standard Deviation | 4.List of CV Scores |
|---|---|---|---|
| RandomForest | 0.8635 | 0.0555 | [0.8196, 0.8189, 0.8163, 0.9284, 0.9344] |
| SVM | 0.8455 | 0.0192 | [0.8366, 0.8264, 0.8276, 0.8714, 0.8656] |
| Decision | 0.7868 | 0.0511 | [0.7518, 0.7401, 0.7437, 0.8481, 0.8504] |
| K-nearest | 0.7287 | 0.0405 | [0.7005, 0.7003, 0.6879, 0.7678, 0.787] |
| GradientBoost | 0.6792 | 0.0049 | [0.6813, 0.6743, 0.6727, 0.6854, 0.6822] |

Table ESM5 Metrics of ML models based on Dataset 3 (1H NMR concise data)

| 1.Algorithm | 2.Accuracy | 3.Precision | 4.Recall | 5.F1 | 6.ROC |
|---|---|---|---|---|---|
| SVM | 0.677 | 0.691 | 0.638 | 0.664 | 0.677 |
| GradientBoost | 0.677 | 0.680 | 0.667 | 0.673 | 0.677 |
| RandomForest | 0.649 | 0.678 | 0.566 | 0.617 | 0.649 |
| Decision | 0.618 | 0.643 | 0.528 | 0.580 | 0.618 |
| K-nearest | 0.613 | 0.669 | 0.445 | 0.535 | 0.613 |

Table ESM6. Five-fold cross-validation for ML based on Dataset 3 (1H NMR concise data)

| 1.Algorithm | 2.Mean CV Score | 3.Standard Deviation | 4.List of CV Scores |
|---|---|---|---|
| RandomForest | 0.7690 | 0.0328 | [0.7483, 0.7433, 0.7357, 0.8071, 0.8108] |
| Decision | 0.7342 | 0.0388 | [0.7058, 0.6995, 0.7027, 0.7773, 0.7858] |
| K-nearest | 0.6891 | 0.0277 | [0.6705, 0.667, 0.6627, 0.717, 0.7282] |
| SVM | 0.6872 | 0.0024 | [0.6895, 0.6905, 0.6849, 0.6859, 0.6849] |
| GradientBoost | 0.6835 | 0.0045 | [0.6923, 0.6813, 0.6804, 0.6832, 0.6802] |

Table ESM7 Metrics of ML models based on Dataset 4 (1H NMR extensive data)

| 1.Algorithm | 2.Accuracy | 3.Precision | 4.Recall | 5.F1 | 6.ROC |
|---|---|---|---|---|---|
| SVM | 0.695 | 0.715 | 0.649 | 0.680 | 0.695 |
| GradientBoost | 0.679 | 0.690 | 0.649 | 0.669 | 0.679 |
| RandomForest | 0.663 | 0.754 | 0.484 | 0.589 | 0.663 |
| K-nearest | 0.602 | 0.661 | 0.419 | 0.513 | 0.602 |
| Decision | 0.595 | 0.615 | 0.506 | 0.555 | 0.595 |

Table ESM8. Five-fold cross-validation for ML based on Dataset 4 (1H NMR extensive data)

| 1.Algorithm | 2.Mean CV Score | 3.Standard Deviation | 4.List of CV Scores |
|---|---|---|---|
| RandomForest | 0.8205 | 0.0548 | [0.7693, 0.7833, 0.775, 0.8851, 0.8896] |
| Decision | 0.7500 | 0.0589 | [0.7048, 0.6998, 0.7015, 0.8166, 0.8273] |
| SVM | 0.7341 | 0.0152 | [0.7205, 0.728, 0.7175, 0.7505, 0.754] |
| K-nearest | 0.6910 | 0.0423 | [0.663, 0.6535, 0.6532, 0.743, 0.7422] |
| GradientBoost | 0.6857 | 0.0078 | [0.672, 0.692, 0.6824, 0.6927, 0.6894] |

Table ESM9 Metrics of ML models based on Dataset 5 (13C NMR concise data & 1H NMR concise data

| 1.Algorithm | 2.Accuracy | 3.Precision | 4.Recall | 5.F1 | 6.ROC |
|---|---|---|---|---|---|
| SVM | 0.778 | 0.796 | 0.747 | 0.771 | 0.778 |
| RandomForest | 0.737 | 0.842 | 0.582 | 0.689 | 0.737 |
| GradientBoost | 0.705 | 0.715 | 0.680 | 0.697 | 0.705 |
| K-nearest | 0.656 | 0.719 | 0.513 | 0.599 | 0.656 |
| Decision | 0.634 | 0.663 | 0.544 | 0.598 | 0.634 |

Table ESM10. Five-fold cross-validation for ML based on Dataset 5 (13C NMR concise data & 1H NMR concise data)

| 1.Algorithm | 2.Mean CV Score | 3.Standard Deviation | 4.List of CV Scores |
|---|---|---|---|
| RandomForest | 0.8608 | 0.0420 | [0.8289, 0.8271, 0.8238, 0.9102, 0.9142] |
| SVM | 0.7921 | 0.0036 | [0.7903, 0.7918, 0.7875, 0.7985, 0.792] |
| Decision | 0.7777 | 0.0461 | [0.7356, 0.7481, 0.7375, 0.8278, 0.8396] |
| K-nearest | 0.7386 | 0.0360 | [0.7103, 0.7133, 0.7045, 0.7868, 0.778] |
| GradientBoost | 0.7276 | 0.0040 | [0.7313, 0.7318, 0.7217, 0.729, 0.724] |

Table ESM11 Metrics of ML models based on Dataset 6 (13C NMR extensive data & 1H NMR concise data)

| 1.Algorithm | 2.Accuracy | 3.Precision | 4.Recall | 5.F1 | 6.ROC |
|---|---|---|---|---|---|
| SVM | 0.764 | 0.790 | 0.719 | 0.753 | 0.764 |
| RandomForest | 0.747 | 0.844 | 0.605 | 0.705 | 0.747 |
| GradientBoost | 0.694 | 0.699 | 0.681 | 0.690 | 0.694 |
| K-nearest | 0.676 | 0.759 | 0.516 | 0.614 | 0.676 |
| Decision | 0.668 | 0.698 | 0.593 | 0.641 | 0.668 |

Table ESM12. Five-fold cross-validation for ML based on Dataset 6 (13C NMR extensive data & 1H NMR concise data)

| 1.Algorithm | 2.Mean CV Score | 3.Standard Deviation | 4.List of CV Scores |
|---|---|---|---|
| RandomForest | 0.8722 | 0.0447 | [0.8356, 0.8339, 0.8376, 0.9284, 0.9254] |
| SVM | 0.7879 | 0.0084 | [0.7806, 0.7866, 0.7775, 0.7995, 0.7953] |
| Decision | 0.7879 | 0.0469 | [0.7498, 0.7561, 0.7435, 0.8448, 0.8453] |
| K-nearest | 0.7496 | 0.0354 | [0.727, 0.7195, 0.7165, 0.786, 0.799] |
| GradientBoost | 0.7103 | 0.0032 | [0.7078, 0.7158, 0.7077, 0.7122, 0.708] |

Table ESM13 Metrics of ML models based on Dataset 7 (13C NMR concise data & 1H NMR extensive data)

| 1.Algorithm | 2.Accuracy | 3.Precision | 4.Recall | 5.F1 | 6.ROC |
|---|---|---|---|---|---|
| SVM | 0.778 | 0.796 | 0.747 | 0.771 | 0.778 |
| RandomForest | 0.714 | 0.828 | 0.540 | 0.654 | 0.714 |
| GradientBoost | 0.711 | 0.722 | 0.687 | 0.704 | 0.711 |
| K-nearest | 0.641 | 0.736 | 0.440 | 0.550 | 0.641 |
| Decision | 0.616 | 0.639 | 0.532 | 0.581 | 0.616 |

Table ESM14. Five-fold cross-validation for ML based on Dataset 7 (13C NMR concise data & 1H NMR extensive data)

| 1.Algorithm | 2.Mean CV Score | 3.Standard Deviation | 4.List of CV Scores |
|---|---|---|---|
| RandomForest | 0.8611 | 0.0464 | [0.8271, 0.8261, 0.8168, 0.9132, 0.9222] |
| SVM | 0.8161 | 0.0176 | [0.8084, 0.8026, 0.7968, 0.8441, 0.8286] |
| Decision | 0.7636 | 0.0570 | [0.7108, 0.7283, 0.7132, 0.8268, 0.8388] |
| GradientBoost | 0.7321 | 0.0107 | [0.724, 0.7308, 0.7177, 0.7412, 0.7467] |
| K-nearest | 0.7275 | 0.0451 | [0.6985, 0.6865, 0.6874, 0.7833, 0.7818] |

Table ESM15 Metrics of ML models based on Dataset 8 (13C NMR extensive data & 1H NMR extensive data)

| 1.Algorithm | 2.Accuracy | 3.Precision | 4.Recall | 5.F1 | 6.ROC |
|---|---|---|---|---|---|
| SVM | 0.792 | 0.824 | 0.742 | 0.781 | 0.792 |
| RandomForest | 0.713 | 0.853 | 0.515 | 0.642 | 0.713 |
| GradientBoost | 0.665 | 0.690 | 0.600 | 0.642 | 0.665 |
| Decision | 0.645 | 0.678 | 0.550 | 0.608 | 0.645 |
| K-nearest | 0.640 | 0.776 | 0.392 | 0.521 | 0.640 |

Table ESM16. Five-fold cross-validation for ML based on Dataset 8 (13C NMR extensive data & 1H NMR extensive data)

| 1.Algorithm | 2.Mean CV Score | 3.Standard Deviation | 4.List of CV Scores |
|---|---|---|---|
| RandomForest | 0.8610 | 0.0591 | [0.8144, 0.8134, 0.8106, 0.9312, 0.9354] |
| SVM | 0.8441 | 0.0195 | [0.8324, 0.8294, 0.8236, 0.8714, 0.8636] |
| Decision | 0.7898 | 0.0502 | [0.7513, 0.7556, 0.7402, 0.8483, 0.8534] |
| K-nearest | 0.7317 | 0.0418 | [0.7, 0.7058, 0.6884, 0.7748, 0.7895] |
| GradientBoost | 0.6809 | 0.0130 | [0.6673, 0.6765, 0.6709, 0.7035, 0.6862] |

Table ESM17 Metrics of ML models based on Dataset 2 (13C NMR extensive data) and molecular features performed without PCA

| | 1.Algorithm | 2.Accuracy | 3.Precision | 4.Recall | 5.F1 | 6.ROC |
|---|---|---|---|---|---|---|
| **2** | RandomForest | 0.830 | 0.880 | 0.764 | 0.818 | 0.830 |
| **3** | GradientBoost | 0.810 | 0.831 | 0.780 | 0.805 | 0.810 |
| **0** | SVM | 0.803 | 0.845 | 0.743 | 0.790 | 0.803 |
| **1** | Decision | 0.765 | 0.794 | 0.715 | 0.753 | 0.765 |
| **4** | K-nearest | 0.673 | 0.725 | 0.558 | 0.631 | 0.673 |

Table ESM18 Five-fold cross-validation for ML based on Dataset 2 (13C NMR extensive data) and molecular features performed without PCA

| 1.Algorithm | 2.Mean CV Score | 3.Standard Deviation | 4.List of CV Scores |
|---|---|---|---|
| RandomForest | 0.8399 | 0.0057 | [0.8311, 0.8464, 0.8397, 0.8454, 0.8367] |
| GradientBoost | 0.8247 | 0.0047 | [0.8162, 0.8297, 0.8262, 0.8277, 0.8238] |
| SVM | 0.7887 | 0.0050 | [0.7831, 0.7956, 0.7915, 0.7903, 0.7828] |
| Decision | 0.7799 | 0.0110 | [0.7752, 0.7962, 0.7744, 0.7885, 0.7654] |
| K-nearest | 0.7687 | 0.0026 | [0.767, 0.7723, 0.7664, 0.7715, 0.7664] |

Table ESM19 Metrics of ML models based on Dataset 2 (13C NMR extensive data) and molecular features performed with PCA

| 1.Algorithm | 2.Accuracy | 3.Precision | 4.Recall | 5.F1 | 6.ROC |
|---|---|---|---|---|---|
| RandomForest | 0.828 | 0.882 | 0.757 | 0.815 | 0.828 |
| GradientBoost | 0.827 | 0.849 | 0.796 | 0.822 | 0.827 |
| Decision | 0.774 | 0.811 | 0.714 | 0.759 | 0.774 |
| K-nearest | 0.677 | 0.691 | 0.641 | 0.665 | 0.677 |
| SVM | 0.508 | 0.545 | 0.101 | 0.171 | 0.508 |

Table ESM20 Five-fold cross-validation for ML based on Dataset 2 (13C NMR extensive data) and molecular features performed with PCA

| 1.Algorithm | 2.Mean CV Score | 3.Standard Deviation | 4.List of CV Scores |
|---|---|---|---|
| RandomForest | 0.9021 | 0.0292 | [0.8762, 0.8779, 0.8832, 0.9247, 0.9484] |
| Decision | 0.8490 | 0.0286 | [0.8211, 0.8299, 0.8291, 0.8697, 0.8951] |
| GradientBoost | 0.8376 | 0.0075 | [0.8284, 0.8314, 0.8371, 0.8494, 0.8416] |
| K-nearest | 0.7329 | 0.0230 | [0.713, 0.716, 0.715, 0.7521, 0.7685] |
| SVM | 0.5139 | 0.0194 | [0.5009, 0.5006, 0.5111, 0.5049, 0.5521] |

Table ESM21 Metrics of ML models based on molecular features only

| 1.Algorithm | 2.Accuracy | 3.Precision | 4.Recall | 5.F1 | 6.ROC |
|---|---|---|---|---|---|
| GradientBoost | 0.807 | 0.826 | 0.778 | 0.801 | 0.807 |
| RandomForest | 0.803 | 0.850 | 0.737 | 0.790 | 0.804 |
| SVM | 0.764 | 0.857 | 0.634 | 0.729 | 0.764 |
| Decision | 0.752 | 0.785 | 0.694 | 0.737 | 0.752 |
| K-nearest | 0.729 | 0.739 | 0.708 | 0.723 | 0.729 |

Table ESM22. Five-fold cross-validation for ML based on molecular features only

| 1.Algorithm | 2.Mean CV Score | 3.Standard Deviation | 4.List of CV Scores |
|---|---|---|---|
| RandomForest | 0.8871 | 0.0286 | [0.866, 0.8676, 0.8593, 0.9132, 0.9296] |
| Decision | 0.8416 | 0.0296 | [0.8183, 0.8165, 0.8189, 0.8679, 0.8863] |
| GradientBoost | 0.8099 | 0.0038 | [0.8053, 0.8165, 0.8082, 0.8087, 0.811] |
| K-nearest | 0.7810 | 0.0221 | [0.7615, 0.7688, 0.7589, 0.8071, 0.8084] |
| SVM | 0.7660 | 0.0031 | [0.7626, 0.7696, 0.767, 0.7686, 0.762] |

Table ESM 23. Classification report of RFC based on IUPAC encoded data

| | precision | recall | f1-score | support |
|---|---|---|---|---|
| Active (target 1) | 0.67 | 0.77 | 0.71 | 2000 |
| Inactive (target 0) | 0.72 | 0.62 | 0.67 | 2000 |
| accuracy | | | 0.69 | 4000 |
| macro avg | 0.70 | 0.69 | 0.69 | 4000 |
| weighted avg | 0.70 | 0.69 | 0.69 | 4000 |

# Figures

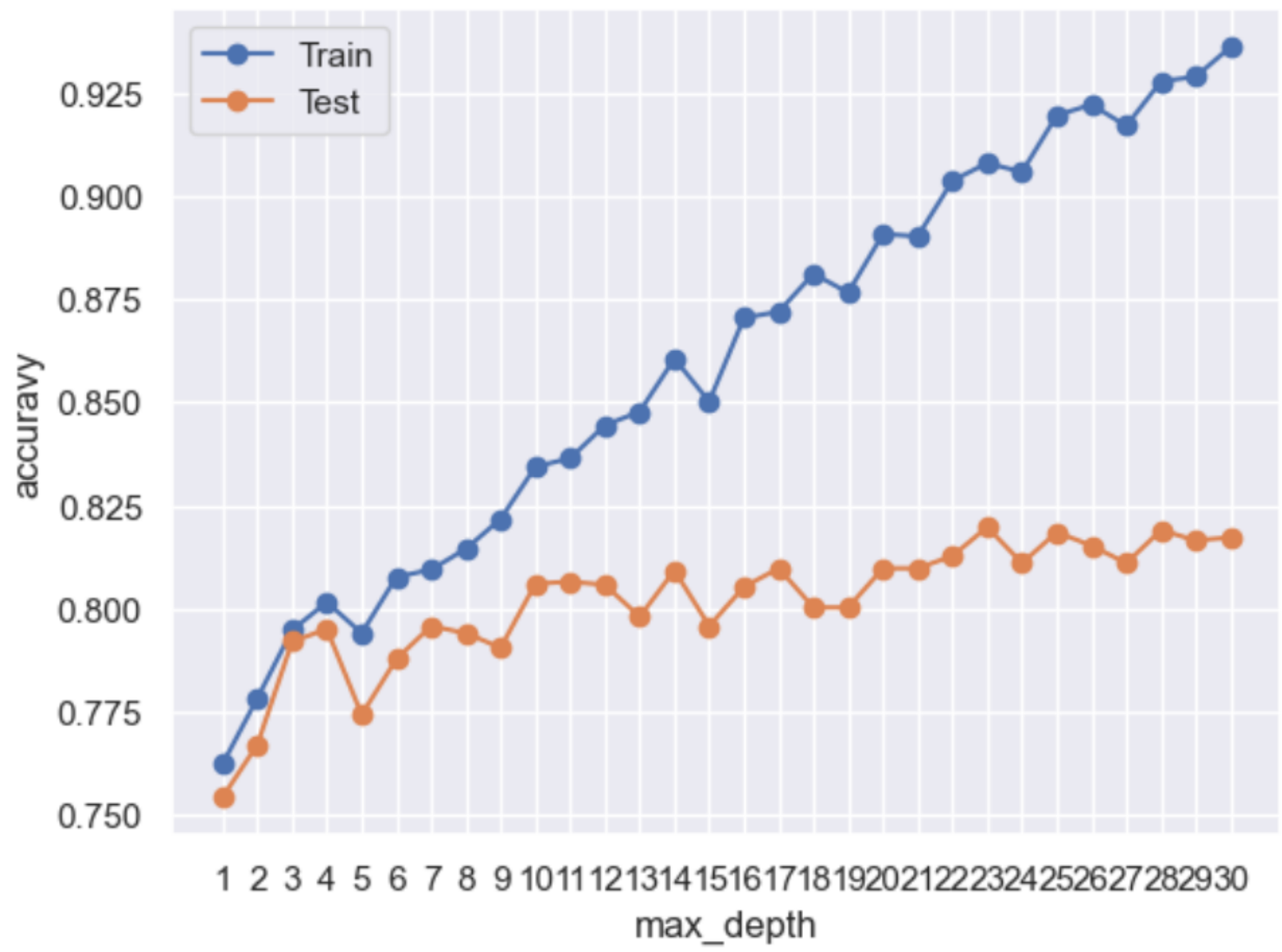

Figure ESM 1. Scrutinising for overfitting of RFC
based on Dataset 2 integrated with molecular feature

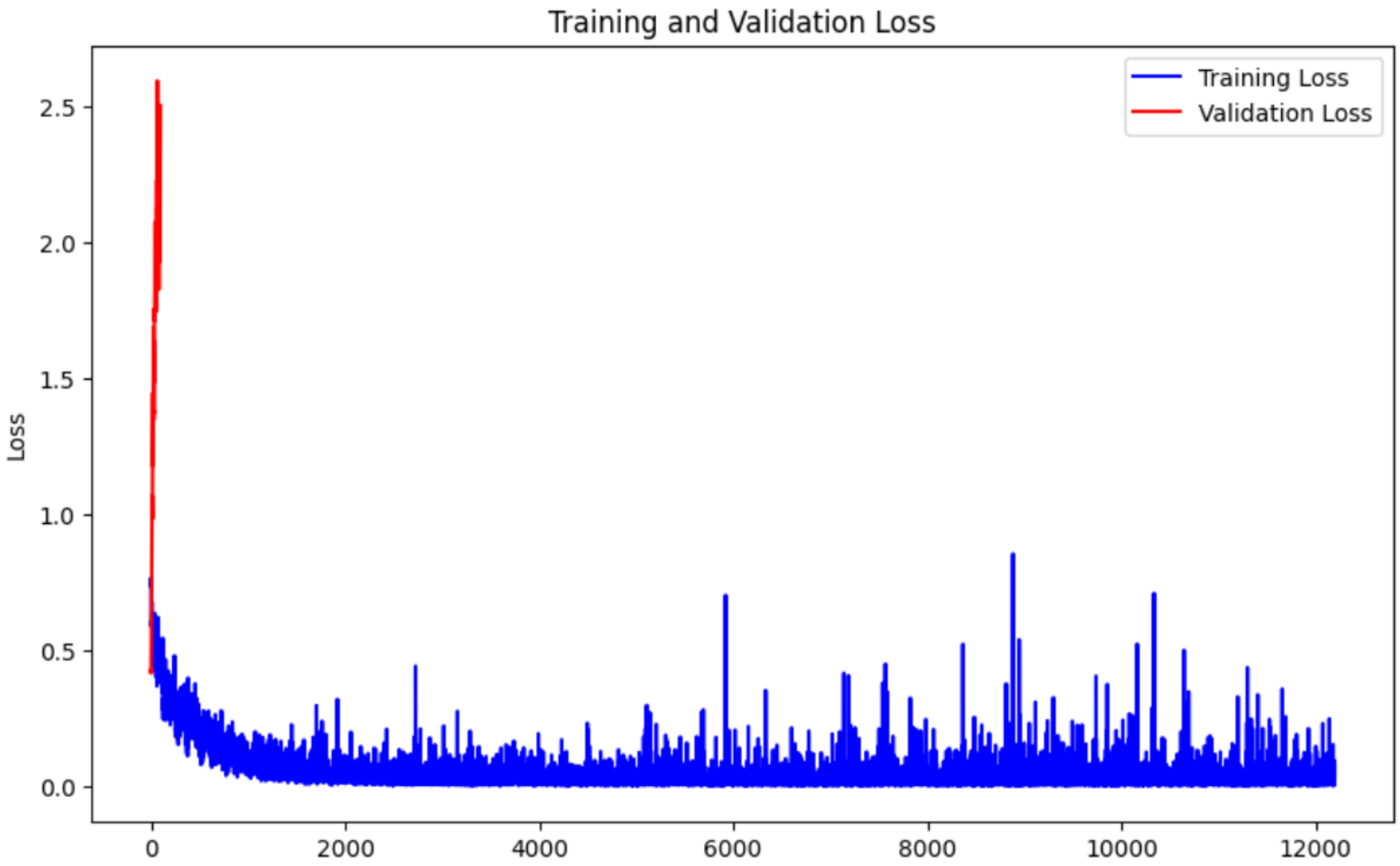

Figure ESM 2. Overfitting check of DNN optimised by Optuna.
Final Training Loss: 0.07168064266443253;
Final Validation Loss: 2.5014856861483667;
Loss Difference (Validation - Training): 2.4298050434839342;
Potential Overfitting Detected.

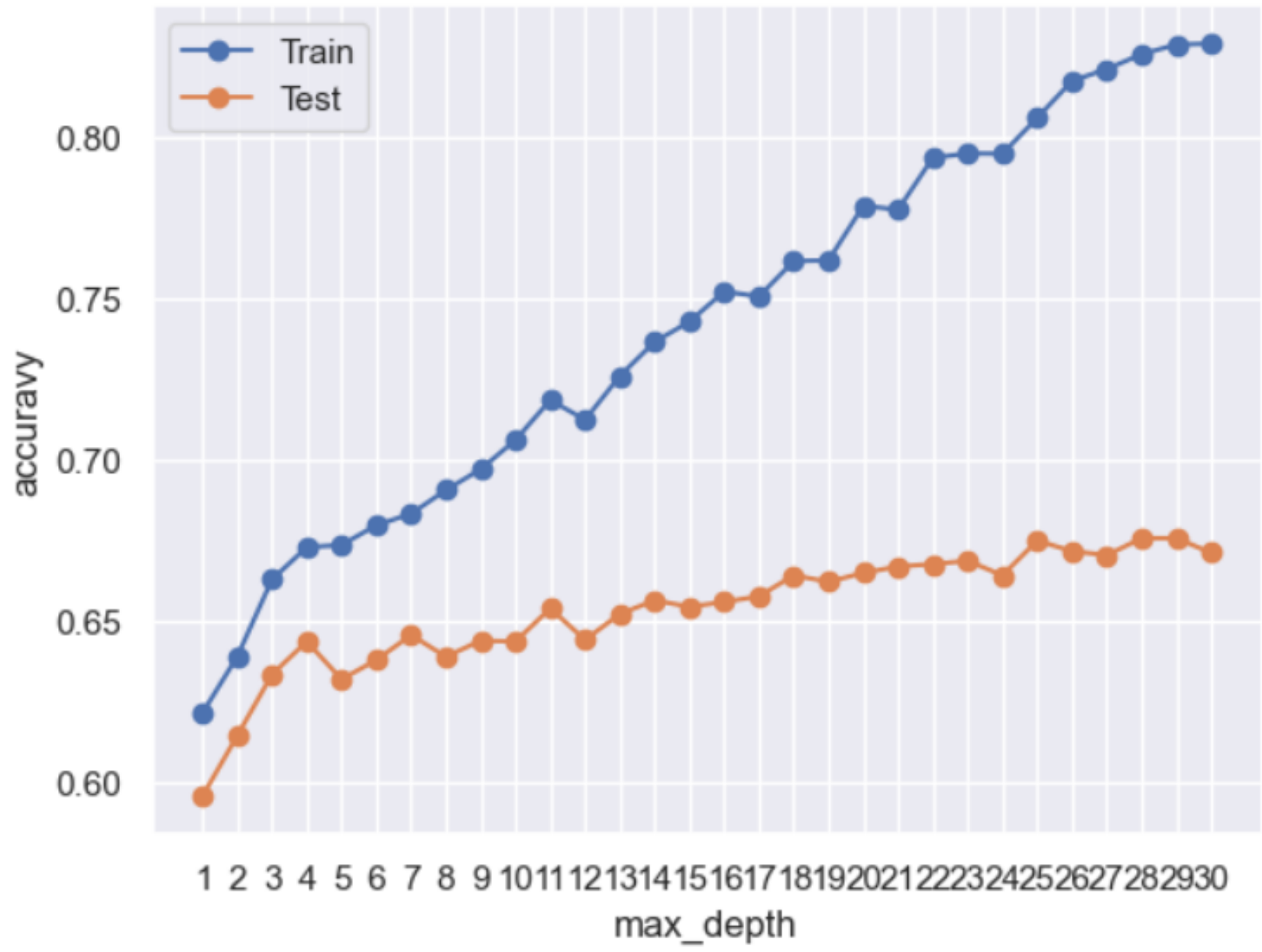

Figure ESM3. Scrutinise for overfitting of RFC based on IUPAC encoded data

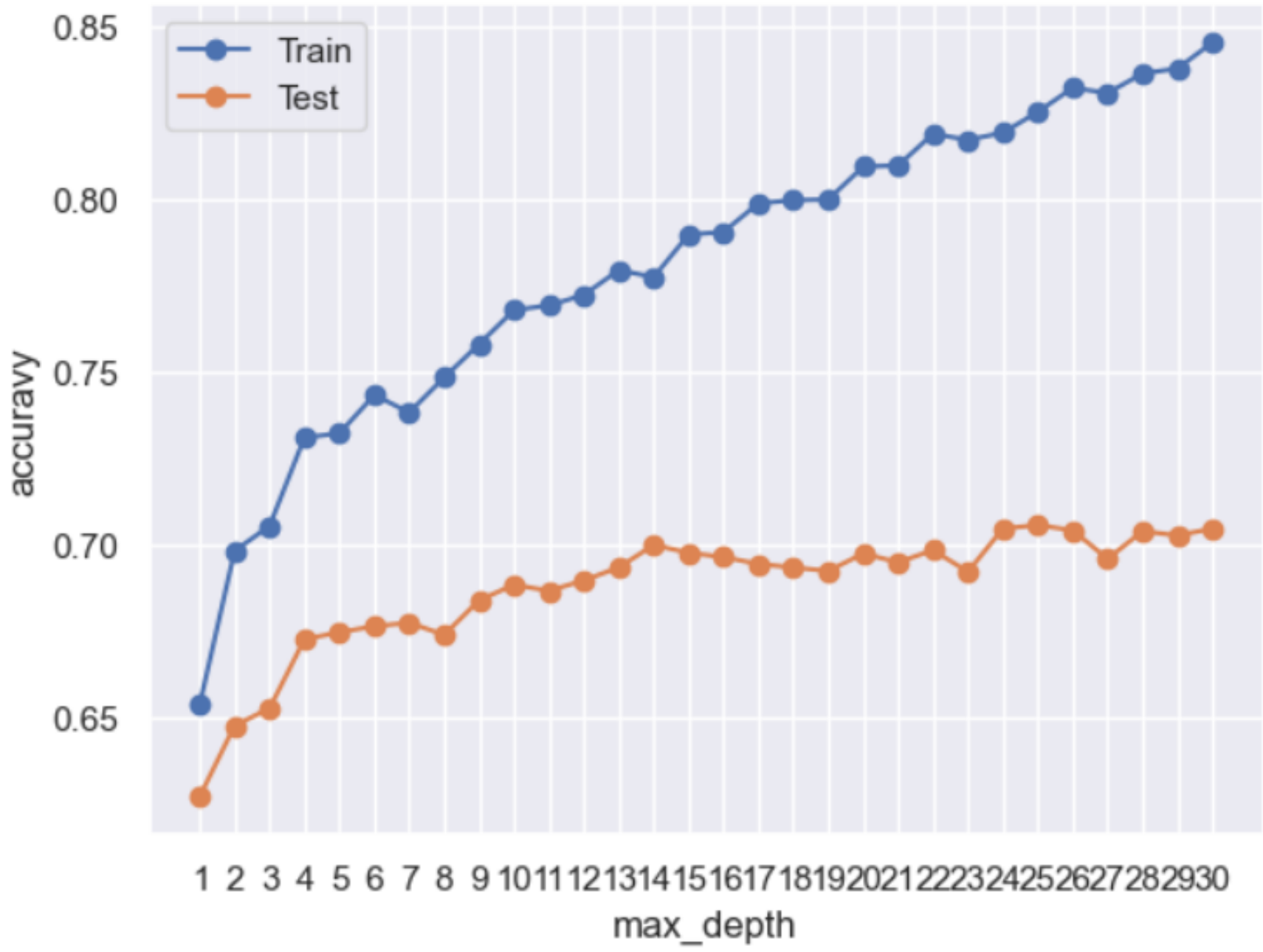

Figure ESM4. Scrutinise for overfitting of RFC
based on IUPAC encoded data hyperparameter tuned by Optuna

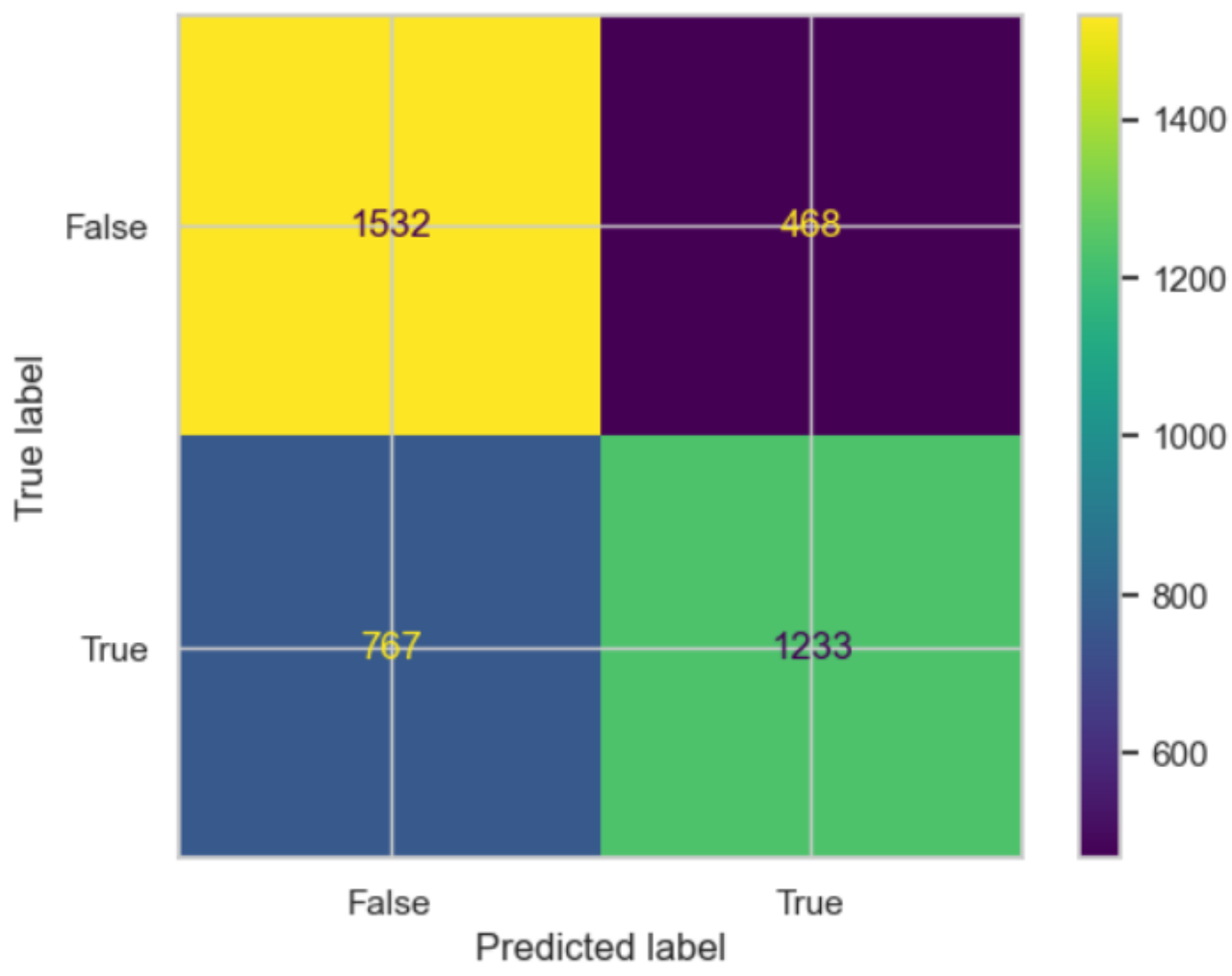

Figure ESM 5. Confusion matrix of ML model based on IUPAC encoded data

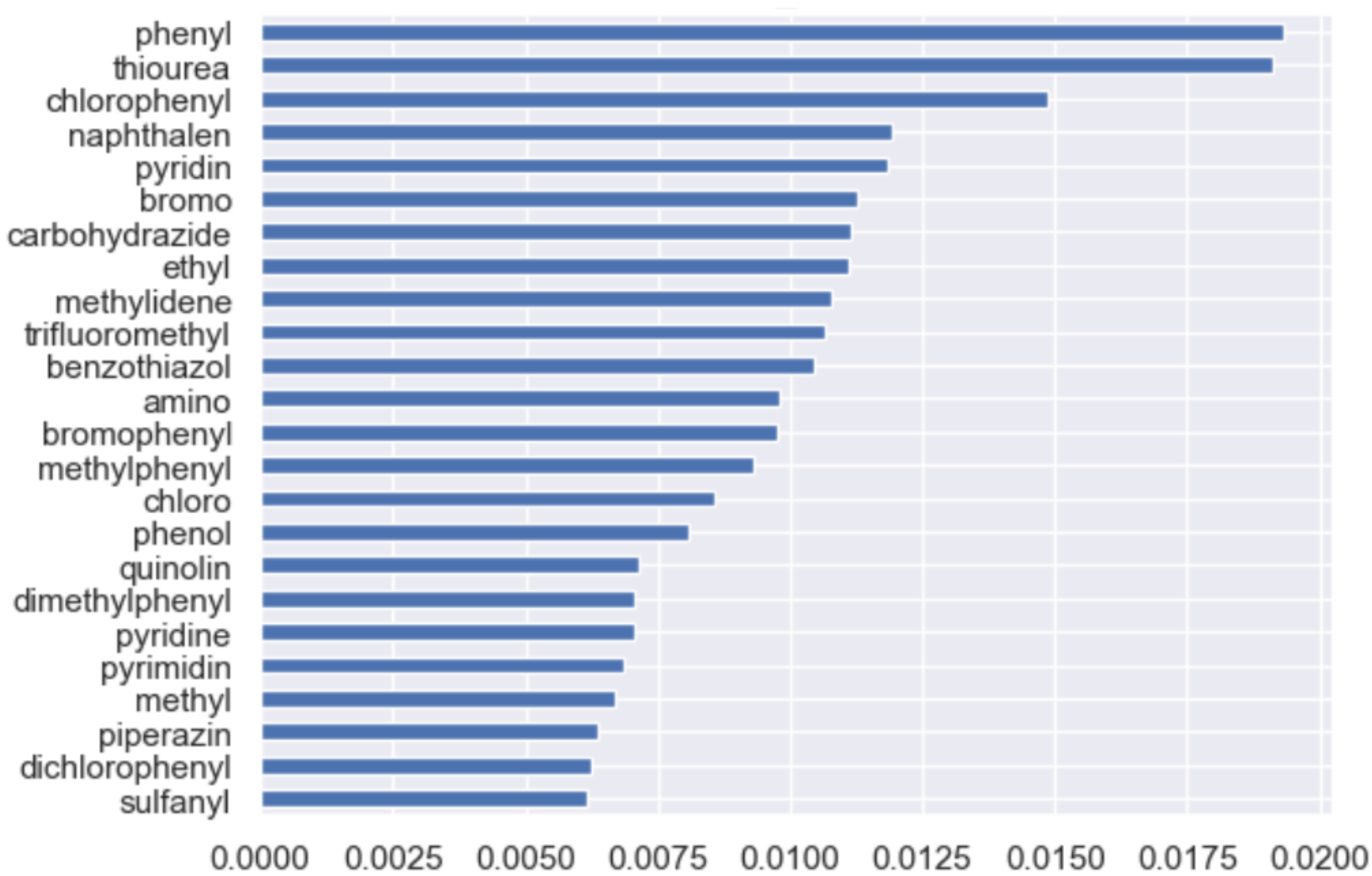

Figure ESM6. Descending order of the functional groups of RFC based on Dataset 2& molecular feature obtained by feature importance algorithm

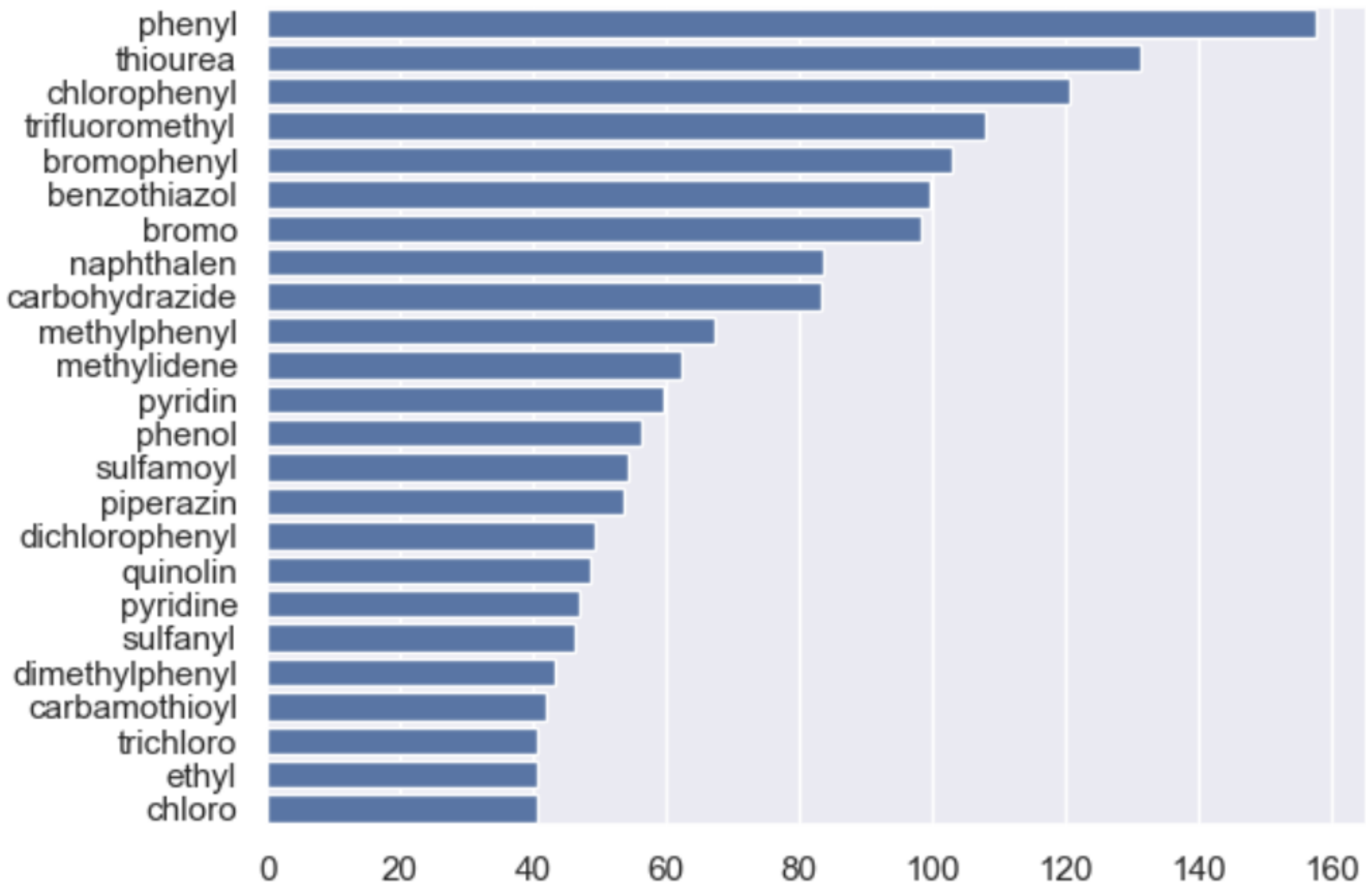

Figure ESM7. Descending order of the functional groups of RFC based on Dataset 2 & molecular features obtained by the permutation importance algorithm